\documentstyle[aps,pre,floats,epsf]{revtex}
\begin{document}
\advance\textheight by 0.2in
\draft



\title{Identification of domain walls in coarsening systems
       at finite temperature}
\author{Haye Hinrichsen and Micka\"el Antoni}
\address{Max-Planck-Institut f\"ur Physik komplexer Systeme,
	 N\"othnitzer Stra{\ss}e 38, D-01187 Dresden, Germany}
\date{January 16, 1998, to appear in Phys. Rev. {\bf E}}

\maketitle

\begin{abstract}
Recently B. Derrida [Phys. Rev. {\bf E 55}, 3705 (1997)]
introduced a numerical technique that allows one to measure the fraction
of persistent spins in a coarsening nonequilibrium system at finite
temperature. In the present work we extend this method in a way that
domain walls can be clearly identified. To this end we consider
three replicas instead of two. As an application we measure the surface
area of coarsening domains in the two-dimensional Ising model at finite
temperature. We also discuss the question of to what extent the
results depend on the algorithmic implementation.
\end{abstract}




\section{Introduction}
\label{Intro}

Dynamical systems quenched from a disordered into an ordered
phase may display interesting coarsening phenomena~\cite{Coarsening}.
A simple example is the Ising model evolving by heat 
bath (HBD) or Glauber dynamics (GD). In the ordering phase of this model, 
starting with random initial conditions, patterns of ferromagnetic
domains are formed whose typical size grows with time as $t^{1/2}$.
For zero temperature, these domains are fully ordered and the domain
walls evolving in time
can be identified as bonds between oppositely oriented
spins~\cite{ZeroTemperature}. For nonzero temperature, however,
it is hard to define domains and domain walls
because it is difficult to distinguish between `true' domains and 
minority islands generated by thermal fluctuations. This 
situation emerges, for example, in the two-dimensional ($2D$) Ising model
at finite temperature below $T_c$ (see Fig.~\ref{FigureOne}a).

Recently B. Derrida~\cite{Derrida} proposed a method that allows one to
measure properties related to coarsening in presence of thermal fluctuations. 
The main idea of this method lies in the comparision of two identical copies
(replicas) $A$ and $B$ of the same system.
Both replicas are submitted to the same thermal noise, i.e., their numerical
updates are determined by the same sequence of random numbers.
Copy~$A$ starts with random initial conditions and begins 
to coarsen whereas copy~$B$ starts from a fully magnetized state
and therefore remains ordered as time evolves. The
assumption is that all spin flips occurring in replica $B$ can be regarded
as thermal fluctuations. Therefore, when a spin flip occurs 
simultaneously in both replicas, it can be considered as a thermal 
fluctuation, otherwise as a fluctuation due to the coarsening process.

In Refs.~\cite{Derrida,Stauffer} this method was used to determine the
fraction of persistent spins \cite{Persistence}
at nonzero temperature as a function of time.
At zero temperature a spin is said to be persistent
up to time $t$ if it never flipped before.
In the Ising model the fraction of persistent spins $r(t)$
decays according to a power law $r(t) \sim t^{-\theta}$ where
$\theta$ is an independent exponent. For the $1d$ Glauber model it was proved
that $\theta=3/8$~\cite{ExactResult} whereas in higher dimensions
$\theta$ could only be determined by numerical simulations~\cite{Persistence}
and approximation methods~\cite{Approximation}. For $T>0$, however,
the fraction of spins that never flipped decays exponentially since thermal
fluctuations occur everywhere at some finite rate. 
To overcome this difficulty, Derrida proposed to consider a
spin as ``persistent'' if its temporal evolution in copies $A$ and $B$
is fully synchronized up to time $t$. Using this definition of
persistence he analyzed the $2D$ Ising model and observed that
$r(t)$ decays algebraically for $0 \leq T \leq T_c$ and
saturates at some finite value for $T>T_c$.
Below the critical temperature the exponent $\theta$ seems
to be the same as for $T=0$
while at $T=T_c$ a different exponent is observed.

An imperfection of the method developed by Derrida is that only one
type of domain can be identified, namely, those that are magnetized in
the same way as system $B$. Therefore different spin flips in
$A$ and~$B$ indicate the presence of oppositely magnetized
domains rather than the presence of a domain wall. This means
that persistent spins can be identified only in those domains 
which have the same orientation as copy~$A$. For the same reason
the method cannot be used to analyze other properties
such as, for example, the dynamics of domain walls.

In the present work we extend Derrida's method in a way that domain
walls can be identified. For this purpose we consider {\em three} replicas
$A,B,C$ instead of two. As before, all replicas are submitted to the
same realization of noise. 
Replica~$A$ starts with random initial conditions and serves
as the master copy in which the coarsening process takes place.
The temporal evolution of replica $A$ is compared with that of 
replicas $B$ and $C$, which start from fully ordered initial
conditions with positive and negative magnetization, respectively. 
As in the original setup, domains with positive magnetization in copy $A$
exhibit the same thermal fluctuations as copy $B$. Likewise, thermal
fluctuations in domains with negative magnetization in copy $A$ are
synchronized with those in copy $C$. Along the domain walls, 
however, fluctuations in replica $A$ may occur that are different from those
in $B$ as well as in $C$. Detecting such fluctuations by an
appropriate observable (to be defined below) we are able
to identify domain walls in a coarsening process at nonzero temperature.
The remarkable efficiency of this method is illustrated in 
Fig.~\ref{FigureOne}b. In addition, our technique allows one
to measure interesting physical quantities such as, for example,
the surface area of domains as a function of time. In what follows
we restrict ourselves to the $2D$ Ising model evolving by HBD
and GD. However, the technique can easily be generalized and
may be applied to many other stochastic coarsening processes.
For example, applying the method to the Potts model with $q>2$ states per
site requires introducing $q+1$ different replicas.

A fundamental problem of numerical methods based on several replicas
evolving under the same noise is that the results may depend on the
algorithmic implementation. This was first observed in so-called
damage spreading (DS) problems. In DS simulations~\cite{DS} two replicas 
of a nonequilibrium system, submitted
to the same thermal noise, are started from slightly different initial
conditions. If the difference between the two copies (the damage) stays
finite or even diverges, the system is said to exhibit damage spreading.
Otherwise, if the two replicas merge into a fully synchronized evolution,
damage is said to heal. Initially DS fascinated researchers, since
it would have indicated the existence of different dynamical
phases in stochastic models
analogous to chaotic and regular phases in deterministic systems.
However, later it was realized~\cite{AlgorithmDep} that such DS phases
are ambiguous since the usage of different but equivalent
algorithms for the same dynamical system can
lead to different DS phase structures~\cite{EytanHayeDK}.
For example, in the Ising model with HBD  
damage always heals while in the case of GD damage may 
spread~\cite{GlauberSpreads}. The reason is that GD and HBD, 
although indistinguishable on a single replica, 
are characterized by different correlations when two or more replicas
are simulated using the same random numbers~\cite{EytanHayeIsing}.
As we are going to demonstrate, a similar algorithmic 
dependence appears in the present numerical technique
where several replicas submitted to the same noise are used to analyze 
coarsening processes. Thus we have to verify to what extent the results
obtained by Derrida~\cite{Derrida} in the case of 
HBD~\footnote{According to the usual terminology 
the dynamical rule used by Derrida in Ref.~\cite{Derrida} is denoted as
{\em heat bath} (spin orienting) dynamics rather than 
Glauber (spin flip) dynamics.} are physically relevant or rather artifacts
of different algorithmic schemes.

The article is organized in the following way. In Sec.~\ref{SectionTwo}
we define HBD and GD as well as an observable by which
domain walls can be detected. Our numerical results for HBD 
are presented in Sec.~\ref{SectionThree}.  By comparing results for
HBD and GD, we address the problem  of algorithmic 
independence in Sec.~\ref{SectionFour}. Finally our results are
summarized and discussed in Sec.~\ref{SectionFive}. 


\section{Detection of domain walls in the Ising model}
\label{SectionTwo}

%
%
\begin{figure}
\epsfxsize=140mm
\centerline{\epsffile{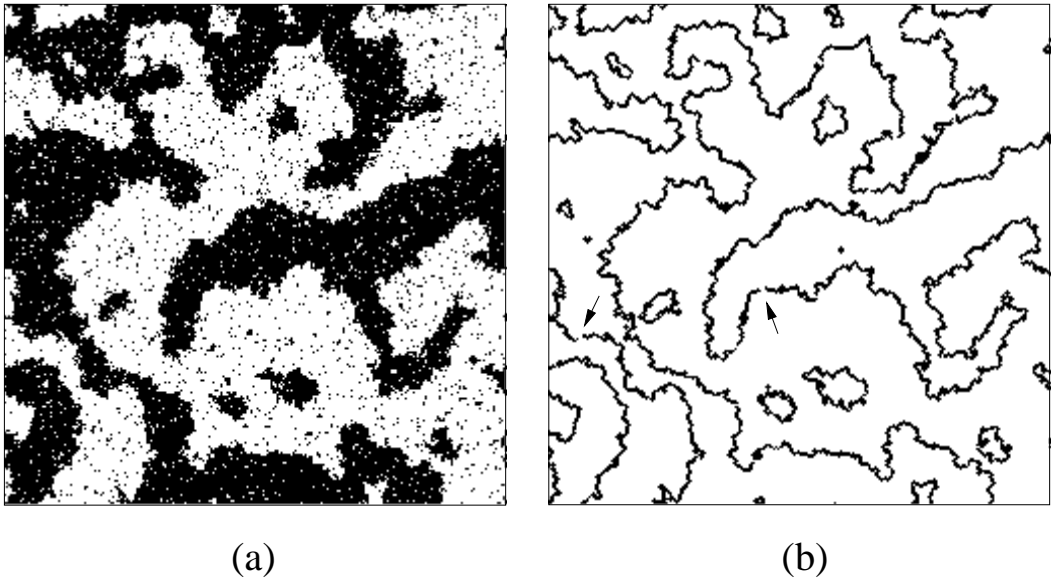}}
\vspace{3mm}
\caption{
The $2D$ Ising model evolving by heat bath dynamics. 
{\bf (a)}:  of a system with $250 \times 250$ sites
and periodic boundary conditions
at temperature $T=0.9 \, T_c$ after $500$ updates.
{\bf (b)}: Corresponding domain walls detected by the observable
$\Delta_i$ defined in Eq.~(\ref{Observable}).
The arrows point at missing links of the contours.
}
\label{FigureOne}
\end{figure}
\noindent

The Ising model evolving by HBD or GD
is defined as follows. Consider a $d$-dimensional
square lattice with spins $\sigma_i(t)=\pm 1$.
The energy at time~$t$ is given by
\begin{equation}
{H} = -\sum_i h_i(t)\sigma_i(t)\,,
\hspace{5mm}
h_i(t) = \sum_j \sigma_j(t), \nonumber
\end{equation}
where $j$ runs over the nearest 
neighbors of site $i$. The local field $h_i(t)$ determines
the transition probability $p_i(t)$ for the spin $\sigma_i$
at time $t$:

\begin{equation}
p_i(t)= \frac{e^{{h_i(t)}/{k_BT}}}
{e^{{h_i(t)}/{k_BT}}+e^{{-h_i(t)}/{k_BT}}}.
\label{eq:pi}
\end{equation}
HBD and GD differ in their update rules:
In HBD the spin $\sigma_i$ is
{\em oriented} according to the local field $h_i(t)$ by
\begin{equation}
\label{HBD}
\sigma_i(t+1)={\rm sgn}[p_i(t)-z_i(t)]\,,
\end{equation}
where $z_i(t)$ are independent random numbers drawn from a 
uniform distribution between 0 and 1. On the other hand,
in GD the spin $\sigma_i$ 
is {\em flipped} depending on its previous orientation:
\begin{equation}
\label{Glauber}
\sigma_i(t+1)=
\left\{ \begin{array}{ll}
\sigma_i(t){\rm sgn}[p_i(t)-z_i(t)] & \mbox{if $ \sigma_i(t)=+1$} \\
\sigma_i(t){\rm sgn}[1-p_i(t)-z_i(t)] & \mbox{if $ \sigma_i(t)=-1$}
\end{array} \right. .
\end{equation}
One can easily verify that in both dynamics
the probability to get $\sigma_i(t+1)=+1$ is the same, as
expected from the equivalence between HBD and GD. 

Let us now consider three replicas $A$, $B$, and $C$ and denote their 
spins by $\sigma^A_i(t)$, $\sigma^B_i(t)$, and $\sigma^C_i(t)$.
As stated before, the initial conditions are
\begin{equation}
\label{InitialConditions}
\sigma^A_i(0) = {\rm sgn}[\frac12-z^{(0)}_i], \qquad
\sigma^B_i(0) = +1, \qquad
\sigma^C_i(0) = -1\,,
\end{equation}
where $z^{(0)}_i$ are random numbers between 0 and 1.
The three replicas evolve under the same realization of noise,
i.e., the same random numbers $z_i(t)$ are used for
the updates of $\sigma^A_i(t)$, $\sigma^B_i(t)$, and $\sigma^C_i(t)$.

We now turn to the observables we want to analyze.
Derrida's definition for the fraction of persistent spins $r(t)$
can be generalized easily:
A spin $\sigma^A_i(t)$ is said to be ``persistent'' up to time $t$
if it experienced exclusively {\em thermal} fluctuations, which means
that it was synchronized either with $\sigma^B_i(t)$ or
with $\sigma^C_i(t)$ for the whole time, i.e.,
\begin{equation}
\label{PersistenceObservable}
r(t) = \frac{1}{N} \sum_{i=1}^N
\left(
\prod_{0 \leq t^\prime \leq t} \frac{1+\sigma^A_i(t)\sigma^B_i(t)}{2} +
\prod_{0 \leq t^\prime \leq t} \frac{1+\sigma^A_i(t)\sigma^C_i(t)}{2}
\right) \,,
\end{equation}
where $N$ is the total number of sites.
We will analyze this quantity numerically
in Secs.~\ref{SectionThree}-\ref{SectionFour}.

In order to identify domain walls, we define an observable 
$\Delta_i(t)$ which compares replicas $A$, $B$, and $C$ at site~$i$
and its nearest neighbors. 
We consider site $i$ as belonging to a domain wall 
(i.e., $\Delta_i(t)=1$) if (a) site~$i$ or at least one of its
nearest neighbors in copies $A$ and $B$
are in different states, {\em and} (b) if site $i$
or at least one of its nearest  neighbors in copies 
$A$ and $C$ are in different states. Since HBD and GD 
evolve independently on two (even and odd) sublattices, we assume that
these nearest neighbors belong to the same sublattice 
(for example, 
if $i=(x,y)$, the nearest neighbors on the same sublattice are
$(x\pm 2,y)$ and $(x,y \pm 2)$).
Formally the observable $\Delta_i(t)$ is defined by
\begin{equation}
\label{Observable}
\Delta_i(t)=
\left(
1-\prod_j\frac{1+\sigma^A_j(t)\sigma_j^B(t)}{2}
\right)
\left(
1-\prod_j\frac{1+\sigma^A_j(t)\sigma_j^C(t)}{2}
\right)\,,
\end{equation}
where $j$ runs over site~$i$ and its nearest neighbors on the same
sublattice. It turns out that this observable allows one to
identify domain walls, as illustrated for HBD in Fig.~\ref{FigureOne}b.

The definition~(\ref{Observable})
appears to be quite complicated since it
involves the nearest neighbors of site $i$. It would have
been more natural to define a local observable
$\delta_i(t)$, which is $1$ if $\sigma^A_i$ is different from
$\sigma^B_i$ and $\sigma^C_i$ (indicating a fluctuation generated
by the coarsening process) and $0$ otherwise:
\begin{equation}
\label{NaiveDelta}
\delta_i(t)=\Bigl(\frac{1-\sigma_i^A(t)\sigma_i^B(t)}{2}\Bigr)
\Bigl(\frac{1-\sigma_i^A(t)\sigma_i^C(t)}{2}\Bigr)\,.
\end{equation}
But, using the initial conditions specified 
in Eq.~(\ref{InitialConditions}), it would turn
out that $\delta_i(t) \equiv 0$ for all $t$ and for both HBD and 
GD. This is due to an overlap of the regions in $A$ 
which are synchronized with either $B$ or $C$.
For HBD this can be proven as follows.
Assume that the three replicas
at time $t$ are in a state where the inequality
\begin{equation}
\label{HBDInequality}
p^C_i(t) \leq p^A_i(t) \leq p^B_i(t)
\end{equation}
holds for all $i$. Since the number of positive spins generated by the
update rule~(\ref{HBD}) for a given random number $z_i(t)$ is
monotonically increasing with $p_i(t)$, one can show that
$h^C_i(t+1) \leq h^A_i(t+1) \leq h^B_i(t+1)$. Therefore the 
inequality~(\ref{HBDInequality}) is also satisfied at the next time step $t+1$.
Since this inequality is satisfied by the
initial conditions~(\ref{InitialConditions}), it
holds by induction at any time. This
implies that events with $\sigma^A_i \neq \sigma^C_i$ and
$\sigma^A_i \neq \sigma^B_i$ do not occur, hence $\delta_i(t) \equiv 0$
for HBD.
In the case of GD the proof is trivial: Since
replicas $B$ and $C$ evolve precisely in opposite states
($\sigma_i^B(t)=-\sigma_i^C(t)$), the observable $\delta_i(t)$ vanishes
automatically. Thus the local observable defined in Eq.~(\ref{NaiveDelta}) 
cannot be used in order to detect domain walls. This is the reason
we use the more complicated definition of Eq.~(\ref{Observable}).


\section{Heat bath dynamics: 
         Persistent spins and the fractal dimension of domain walls}
\label{SectionThree}

\begin{figure}
\epsfxsize=160mm
\centerline{\epsffile{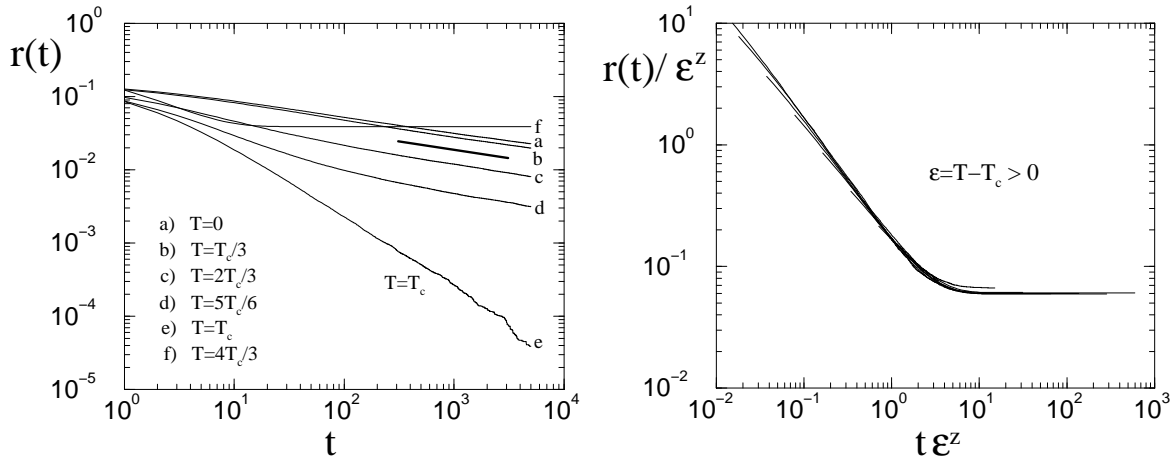}}
\caption{
Heat bath dynamics. Measurement of the fraction of persistent
spins $r(t)$  at various temperatures as a function of time
(c.f. Ref.~[3]). 
The bold line indicates the slope $-0.22$.
The right hand graph shows a data collapse of the rescaled quantity 
$r(t)/\epsilon^2$ in the supercritical regime $T>T_c$.
}
\label{FigureTwo}
\end{figure}

\begin{figure}
\epsfxsize=160mm
\centerline{\epsffile{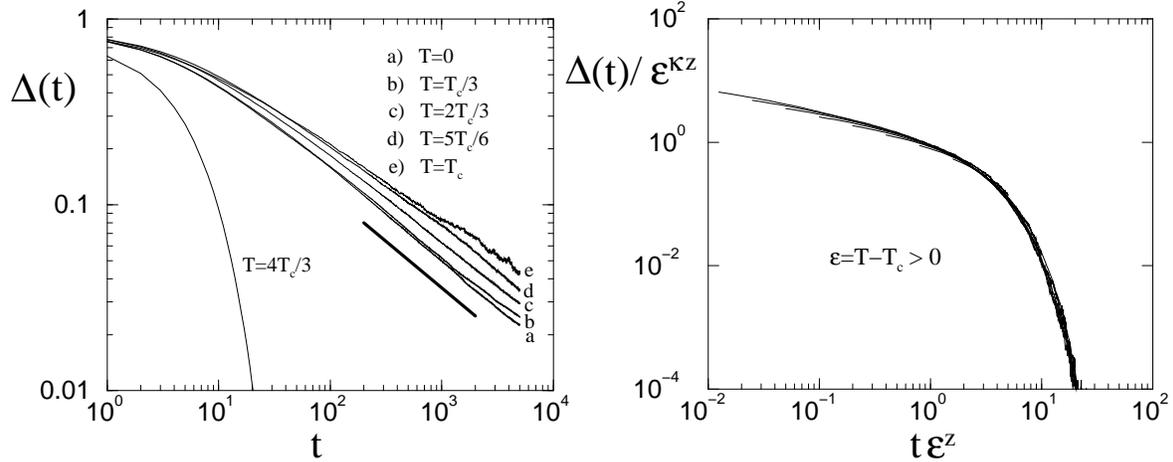}}
\caption{
Heat bath dynamics. Measurement of the total Peierls length~$\Delta(t)$ 
at various temperatures as a function of time. The bold
line indicates slope $-0.5$. The right hand graph 
shows the corresponding data collapse of the 
rescaled Peierls length $\Delta(t)/\epsilon$ in the supercritical regime.
}
\label{FigureThree}
\end{figure}

In this section we present numerical results for the
$2D$ Ising model with HBD. We simulate three replicas of
a system of size $1000 \times 1000$ with periodic boundary
conditions. Starting with the initial conditions~(\ref{InitialConditions})
we measure the fraction of persistent spins $r(t)$
as defined in Eq.~(\ref{PersistenceObservable}) and the average
Peierls length (circumference) of the domains
\begin{equation}
\Delta(t) = \frac{1}{N} \, \sum_i \Delta_i(t)\,.
\end{equation}
The quantities $r(t)$ and $\Delta(t)$ are measured up to
$5000$ time steps and averaged over $10$ independent runs.
Our simulation data are shown in Figs.~\ref{FigureTwo}-\ref{FigureThree}.

%
%
\noindent
%
%

%
%
\noindent

The results for the fraction of persistent spins $r(t)$ 
(see left hand graph in Fig.~\ref{FigureTwo}) are in full
agreement with Ref.~\cite{Derrida}. For $T=0$ we observe an algebraic
decay $r(t) \sim t^{-\theta}$ with $\theta=0.23 \pm 0.03$. For $0 \leq T<T_c$ 
$r(t)$ decreases rapidly on short time ranges while for longer times it
crosses over to a $T$-independent power law decay. Precisely at the 
critical temperature, however, $r(t)$ seems to vanish like
$r(t) \sim t^{-\theta_c}$ with a different exponent $\theta_c \simeq 1$
(the physical relevance of this exponent will be discussed in the
Sec.~\ref{SectionFour}). Finally, for $T>T_c$, $r(t)$ saturates at some
finite value. This can be explained as follows. For $T>T_c$ the total
magnetization in copies $B$ and $C$ decays exponentially. It has been
shown~\cite{GlauberSpreads} that under these conditions any difference between 
two replicas evolving by HBD vanishes exponentially, i.e., 
damage heals spontaneously. This means that the replicas $A,B,C$ converge
and eventually merge into a fully synchronized evolution within
finite time and consequently a finite fraction of persistent spins 
survives. Notice that in the limit $T \rightarrow \infty$ 
all replicas are already synchronized after a single time step. For finite
temperatures we observe a scaling behavior (see right hand graph
in Fig.~\ref{FigureTwo})
\begin{equation}
\label{FirstScalingEquation}
r(t) \sim (T-T_c)^z\,f(t(T-T_c)^z) \,,
\qquad \qquad (T>T_c)
\end{equation}
where $z\simeq 2.125$ is the dynamical critical exponent of
HB dynamics and $f(x)$ is a scaling function which
 behaves as $f(x) \sim 1/x$ for 
$x\rightarrow 0$ and saturates for $x \rightarrow \infty$.\\

The results for the total Peierls length of the domains $\Delta(t)$ 
illustrated in Fig.~\ref{FigureThree} 
indicate a power law behavior $\Delta(t)=t^{-\kappa}$
with $\kappa \simeq 0.5$ in the regime $0 \leq T < T_c$ 
and an exponential decay in the disordered phase $T > T_c$
(to determine the behavior at $T=T_c$ more numerical 
effort would be needed, c.f. Ref.~\cite{CueilleFuture})
that can be explained by the synchronization of 
the copies within finite time. The corresponding
scaling behavior reads
\begin{equation}
\label{SecondScalingEquation}
\Delta(t) \sim (T-T_c)^{\kappa z}\,g(t(T-T_c)^z) \,,
\qquad \qquad (T>T_c)
\end{equation}
where $g(x)$ is a scaling function as shown in the 
right hand graph in Fig.~\ref{FigureThree}.

\begin{figure}
\epsfxsize=120mm
\centerline{\epsffile{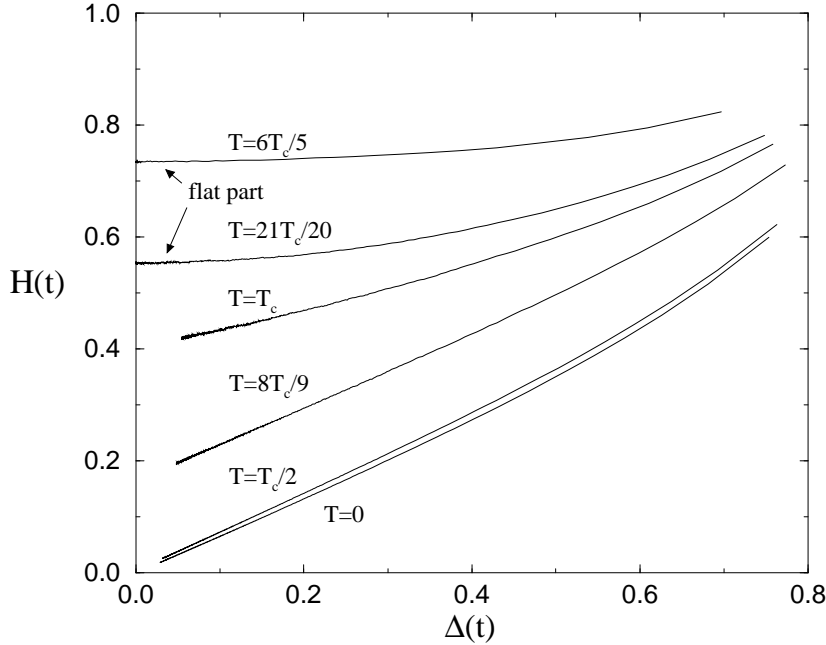}}
\caption{
Total energy ${H}(t)$ of a coarsening Ising model as a function of the 
Peierls length $\Delta(t)$ for various temperatures. 
}
\label{FigureFour}
\end{figure}

The exponent $\kappa \simeq 0.5$ in the subcritical regime may be interpreted
as follows. The total number of domains $n(t)$ in a large
but finite sized system decreases as $n(t) \sim t^{-d/2}$. 
Moreover, the average size of the domains $\xi$ grows with
time as $\xi \sim t^{1/2}$. Hence in
the $2D$ Ising model the Peierls length behaves as
\begin{equation}
\label{SurfaceArea}
n(t) \xi \sim \Delta(t) \sim t^{-1/2}\,.
\end{equation}
This result suggests that surfaces of domains in a coarsening
Ising model are {\em regular}, i.e., they do not have a fractal structure
at finite temperatures $T \leq T_c$. This result can be explained as
follows. The coarsening process is driven by the tendency of the systen
to minimize its energy ${H}$, i.e., to minimize the surface area
(the Peierls length $\Delta(t)$) of the domains. This makes it highly
unlikely for the domain walls to form fractal structures. This mechanism
works not only at zero temperature but prevails in the entire subcritical 
regime. To support this argument we plotted the energy ${H}$ against
the Peierls length in Fig.~\ref{FigureFour}. For $T<T_c$ the curves are
monotonically decreasing with time and seem to have a well defined minimum at
$\Delta=0$. This suggests that the mechanism for coarsening, apart
from different time scales, is the same in the whole subcritical regime.
In the disordered phase, however, the curves have a flat shape close
to $\Delta=0$ and therefore the dynamics of $\Delta(t)$ is no longer
driven by the minimization of energy.

%
%
\noindent
%
%


\section{The problem of algorithmic dependence}
\label{SectionFour}
As outlined in the Introduction, any numerical technique based on
several replicas of a nonequilibrium system evolving under the same
realization of noise may depend in a crucial way on 
the algorithmic implementation of the dynamics. We now discuss
this dependence in the present problem by comparing HBD and GD.

HBD and GD are two different but equally legitimate algorithmic 
implementations of the {\em same} nonequilibrium process that 
mimics the evolution of an Ising model in contact with a thermal reservoir.
To understand this, it may be helpful to rewrite the update rule
for GD~(\ref{Glauber}) as
\begin{equation}
\label{RewrittenGlauber}
\sigma_i(t+1)=
\left\{ \begin{array}{ll}
{\rm sgn}[p_i(t)-z_i(t)] & \mbox{if $ \sigma_i(t)=+1$} \\
{\rm sgn}[p_i(t)-\{1-z_i(t)\}] & \mbox{if $ \sigma_i(t)=-1$}
\end{array} \right. .
\end{equation}
This rule differs from HBD only inasmuch -- depending on 
$\sigma_i(t)$ -- the random number $z_i(t)$ or $1-z_i(t)$ is used.
Since in a simulation of a single replica each random number is 
used only once, it makes no difference whether $z_i(t)$ or $1-z_i(t)$
enters the update rule and therefore both dynamical procedures
are fully equivalent. In other words, looking at the 
space-time trajectory of a {\em single} Ising model,
one cannot distinguish whether it was generated by HBD or GD.
However, if we consider more than one replica evolving under the same
noise, each random number $z_i(t)$ is used several times leading to
different correlations $\em between$ the replicas depending on whether
HBD or GD are used. 

HBD and GD are two members of an infinite family of
equivalent dynamical rules~\cite{InterpolGlauberHBD,EytanHayeIsing}. 
All these rules are equally legitimate and there is no reason
to prefer particular rules such as HBD or GD. {\em Physical}
properties, however, should not depend on the choice of the dynamical 
procedure. Thus, in order to prove the algorithmic independence of a 
specific result, one would have to verify all these rules separately.
Since this is practically impossible, we will restrict ourselves to
the examples of HBD and GD. We will show that some of the previous results 
are affected by a change of the procedure and hence are 
physically irrelevant whereas others are not.

%
%
\begin{figure}
\epsfxsize=140mm
\centerline{\epsffile{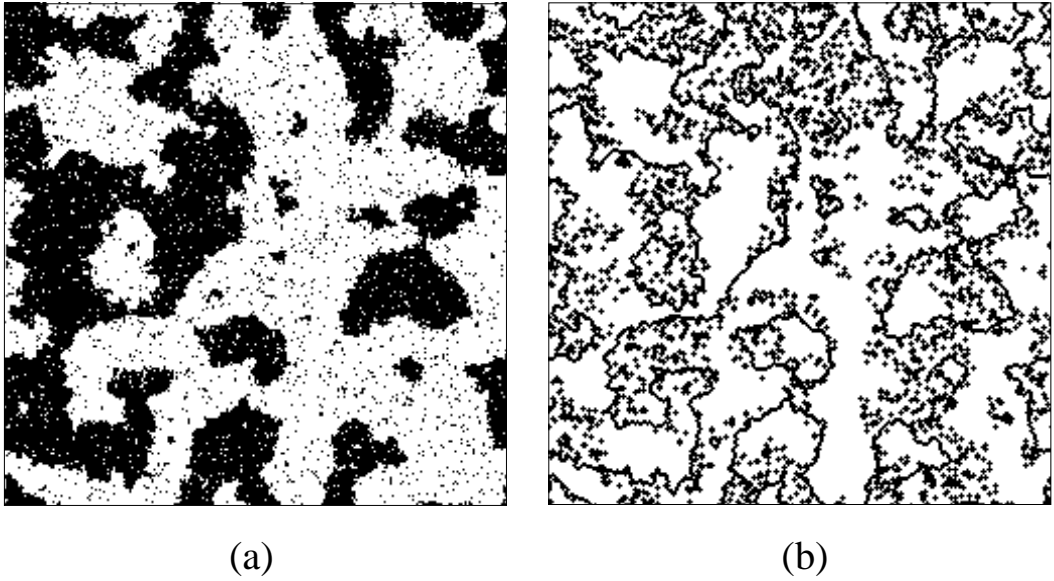}}
\vspace{3mm}
\caption{
The $2D$ Ising model evolving by Glauber dynamics.
{\bf (a)}: Snapshot of a system with $250 \times 250$ sites
and periodic boundary conditions
at temperature $T=0.9 \, T_c$ after $500$ time steps.
{\bf (b)}: Corresponding domain walls detected by the
observable $\Delta_i$. Notice that all contours
are strictly closed.
}
\label{FigureFive}
\end{figure}
\noindent

We now repeat the numerical simulations described in the previous section
using GD instead of HBD. A snapshot of the simulation 
is shown in Fig.~\ref{FigureFive}. Comparing Figs.~\ref{FigureOne}b
and~\ref{FigureFive}b we notice that the HBD algorithm yields very clean
shapes for the Peierls contours while GD produces a lot of additional
fluctuations. On the other hand the  contours in the Glauber case 
are strictly closed because of symmetry reasons 
whereas for HBD there are missing links
(two of them are marked by arrows in Fig.~\ref{FigureOne}b).

We would like to note that any dynamical rule used in the present problem
should synchronize the evolution of replicas $A,B$ or $A,C$ in the
interior of large domains. In the terminology of DS this means that damage 
has to heal. As HBD is the most correlated algorithm, damage
heals very rapidly and thus yields high resolution in the
determination of the walls.
On the other hand GD in $2D$ is known to exhibit a DS transition 
at $T=T_s\simeq 0.95 \ T_c$~\cite{GrassbergerTsGlauber}. The simulations 
in Fig.~\ref{FigureFive}b at $T=0.9\,T_c$ take place very close to this 
transition which explains why the output is rather noisy.

%
%
\begin{figure}
\epsfxsize=160mm
\centerline{\epsffile{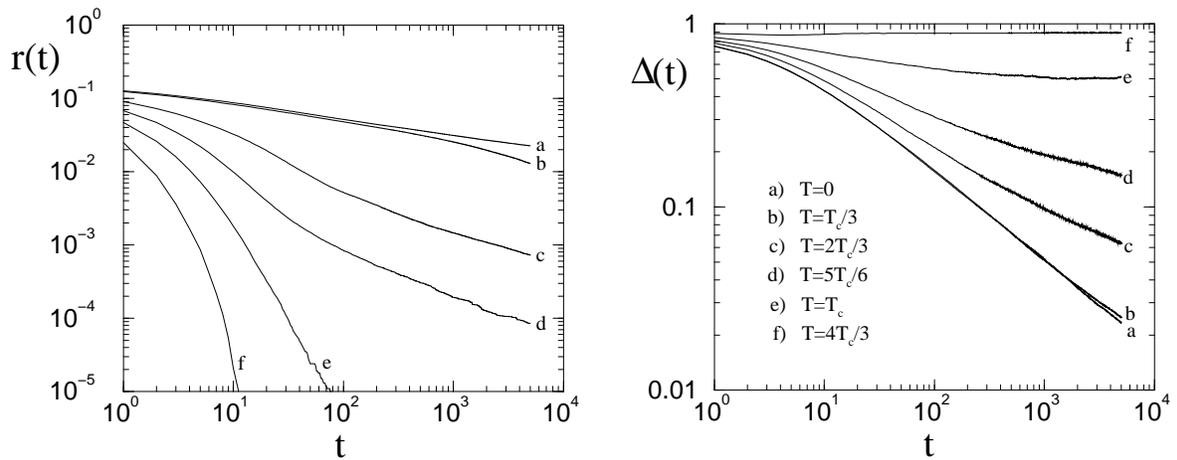}}
\caption{
Glauber dynamics. Measurement of the fraction of persistent
spins $r(t)$ and the total Peierls length of domain walls $\Delta(t)$
as a function of time.
}
\label{FigureSix}
\end{figure}
\noindent

The numerical results for GD and HBD are
significantly different (see Fig.~\ref{FigureSix}). Only at zero 
temperature where GD and HBD coincide does one obtain 
identical results. For $0 < T < T_c$ the fraction of persistent spins $r(t)$ 
exhibits similar behavior in both cases: It first decays 
and then crosses over to an algebraic decay $r(t)\sim t^{-\theta}$
with the same $\theta \simeq 0.2$ as in the $T=0$ case. This 
suggests that the crossover phenomenon is an intrinsic physical property
of the $2D$ Ising model rather than an algorithmic artifact. 
At criticality, however, GD indicates an algebraic decay $r(t) \sim t^{-2}$.
This is clearly different from the decay $r(t) \sim t^{-1}$ observed in the
case of HBD. Therefore one may wonder about the physical meaning
of $r(t)$ at criticality. This problem may be circumvented by a systematic
study of the crossover time as a function of temperature.

The results for the total Peierls length of the domains are even more
contradictory. While in the HBD case we found an algebraic decay
$\Delta(t) \sim t^{-1/2}$ for $0 \leq T \leq T_c$, we observe 
continuously varying exponents in the case of GD. If
this was true, it would imply that the Peierls length per domain grows
like $\Delta(t)/n(t) \sim \xi^\kappa$ with some temperature-dependent
exponent $\kappa>1$ indicating a ``fractal'' structure of the domain walls.
Indeed, visual inspection of Fig.~\ref{FigureFive}b makes it intuitively clear
how such a ``fractal'' structure emerges. It is therefore tempting to
discard the results for GD and to declare the smooth
lines in Fig.~\ref{FigureOne}b as `true' domain walls. However, as
explained above, we have no justification for doing so! Therefore the method
described in this paper cannot be used to determine the fractal 
dimension of domain walls on a safe ground. Nevertheless HBD 
gives us a lower bound for the exponent~$\kappa$. It is thus very likely,
although not strictly proven, that domain walls in the coarsening
Ising model at nonzero temperature do not have a fractal structure.


\section{Concluding remarks}
\label{SectionFive}
By introducing three replicas of a $2D$ Ising model
evolving under the same realization
of noise we extend the numerical method proposed by 
B. Derrida~\cite{Derrida} in 
a way that domain walls can be identified. 
Using HBD we measured the Peierls length of coarsening
domains in the $2D$ Ising model. Our simulations confirm previous results and
suggest that domain walls in the Ising model at $T<T_c$ are regular, 
i.e., they do not have fractal properties.

A fundamental problem appearing here and related to the use of several replicas
is the algorithmic dependence. As an example we compared HBD and GD. 
It turns out that the persistence exponent below the critical 
temperature is the same in both cases which suggests that this 
result is in fact related to  a physical property rather than 
algorithmic artifacts. On the other hand, the Peierls length of 
the domain walls grows differently for HBD and GD, 
although it seems that HBD gives the correct result.

In this context we would like to note that very recently a different method
for the numerical estimation of the persistence exponent $\theta$ has been
proposed~\cite{Cueille} where the persistence probability is defined
in terms of spin blocks. By analyzing the scaling behavior for different 
block sizes one can determine $\theta$ even at finite temperature. Although 
this method cannot be used to identify domain walls, it is very interesting 
since it uses only a single replica wherefore the results do not depend 
on whether HBD or GD is used. In agreement with 
Ref.~\cite{Derrida} and the present work the authors observe that the 
exponent $\theta$ is the same in the entire subcritical regime.  
It would be interesting to use this method in order to determine~$\theta$
at criticality~\cite{CueilleFuture}. 

\vspace{3mm}
\noindent
Acknowledgments:
We thank E. Domany, P. Grassberger, 
G. Schliecker, C. Sire and D. Stauffer for valuable 
discussions and comments. We also thank S. Cueille for
pointing out to us the correct dynamical exponent in
Eqs.~(\ref{FirstScalingEquation})-(\ref{SecondScalingEquation}).


\begin{thebibliography}{99}

\bibitem{Coarsening}	P. C. Hohenberg and B. I. Halperin,
			Rev. Mod. Phys. {\bf 49}, 435 (1977);
			A. J. Bray, Adv. Phys. {\bf 43}, 357 (1994).

\bibitem{ZeroTemperature}
			A. J. Bray, J. Phys. {\bf A 23}, L67 (1990);
			J. G. Amar and F. Family, 
			Phys. Rev. {\bf A 41}, 3258 (1990);
			B. Derrida and R. Zeitak, 
			Phys. Rev. {\bf E 54}, 2513 (1996).

\bibitem{Derrida} 	B. Derrida, Phys. Rev. {\bf E 55}, 3705 (1997).

\bibitem{Stauffer}	D. Stauffer, Int. J. Mod. Phys. {\bf C 8}, 361 (1997).

\bibitem{Persistence}	B. Derrida, A. J. Bray, and C. Godr\`eche,
			J. Phys. {\bf A 27}, L357 (1994);
			D. Stauffer, J. Phys. {\bf A 27}, 5029 (1994).
			
\bibitem{ExactResult}	B. Derrida, V. Hakim, and V. Pasquier,
			Phys. Rev. Lett. {\bf 75}, 751 (1995).
			
\bibitem{Approximation} S. N. Majumdar and C. Sire, 
			Phys. Rev. Lett. {\bf 77}, 1420 (1996);
			S. N. Majumdar, C. Sire, A. J. Bray, and S. J. Cornell,
			Phys. Rev. Lett. {\bf 77}, 2867 (1996);
			B. Derrida, Phys. Rev. Lett. {\bf 77}, 2871 (1996).

\bibitem{DS}		S. A. Kaufmann, 
			J. Theor. Biol. {\bf 22}, 437 (1969);
			M. Creutz, Ann. Phys. {\bf 167}, 62 (1986);
			B. Derrida and G. Weisbuch,
			Europhys. Lett. {\bf 4}, 657 (1987).

\bibitem{AlgorithmDep}	P. Grassberger, J. Stat. Phys. {\bf 79}, 13 (1995).

\bibitem{EytanHayeDK} 	H. Hinrichsen, J. S. Weitz and E. Domany,
                   	J. Stat. Phys. {\bf 88}, 617 (1997).

\bibitem{GlauberSpreads}
			H. Stanley, D. Stauffer, J. Kert\'esz and H. Herrmann,
			Phys. Rev. Lett. {\bf 59}, 2326 (1987);
			A. M. Mariz, H. J. Herrmann and L. de Arcangelis, 
			J. Stat. Phys. {\bf 59}, 1043 (1990).

\bibitem{EytanHayeIsing} 
			H. Hinrichsen and E. Domany, 
			Phys. Rev. {\bf E 56}, 94 (1997).

\bibitem{InterpolGlauberHBD}
			A. M. Mariz and H. J. Herrmann,
			J. Phys. {\bf A 22}, L1081 (1989).

\bibitem{GrassbergerTsGlauber}
			P. Grassberger, J. Phys. {\bf A 28}, L67 (1995);
			T. Vojta, J. Phys. {\bf A 30}, L7 (1997).
			
\bibitem{Cueille}	S. Cueille and C. Sire, unpublished	
			(preprint cond-mat/9707287).

\bibitem{CueilleFuture}	S. Cueille and C. Sire, in preparation.

\end{thebibliography}
\end{document}